\documentclass[]{statsoc}
\usepackage{enumitem}
\usepackage{natbib}
\usepackage{setspace}
\usepackage{geometry}                
\usepackage{graphicx}
\usepackage{amssymb}
\usepackage{amsmath}
\usepackage{epstopdf}
\usepackage[linesnumbered,vlined,algoruled]{algorithm2e}
\usepackage{url}
\usepackage{mathtools}
\DeclarePairedDelimiter{\ceil}{\lceil}{\rceil}

\title[Accurate Quantiles Using $t$-Digests]{Computing Extremely Accurate Quantiles Using $t$-Digests}

\author[Dunning \it{et al.}]{Ted Dunning\thanks{Corresponding author}}
\address{MapR Technologies, Inc \\ Santa Clara, CA}
\email{ted.dunning@gmail.com}
\author[Dunning and Ertl]{Otmar Ertl}
\address{Dynatrace\\Linz, Austria}

\email{otmar.ertl@gmail.com}
\date{}                                           
\begin{document}
\begin{abstract}
We present on-line algorithms for computing approximations of rank-based statistics that give high accuracy, particularly near the tails of a distribution, with very small sketches.  Notably, the method allows a quantile $q$ to be computed with an accuracy relative to $\max(q, 1-q)$ rather than absolute accuracy as with most other methods.  This new algorithm is robust with respect to skewed distributions or ordered datasets and allows separately computed summaries to be combined with no loss in accuracy.

An open-source Java implementation of this algorithm is available from the author. Independent implementations in Go and Python are also available.

\end{abstract}
\maketitle
\section{Introduction}
Given a set of numbers, it is often desirable to compute rank-based statistics such as the median, 95-th percentile or trimmed means in an on-line fashion without having to retain all of the samples or to wait until all data has been processed. In many cases, there is an additional requirement that only a small data structure needs to be kept in memory as data is processed in a streaming fashion.  Traditionally, such statistics have been computed by sorting the data and then either finding the quantile of interest by interpolation or by re-processing all samples within particular quantile ranges.  This sorting approach can be infeasible for very large datasets or when quantiles of many subsets must be calculated. This infeasibility has led to interest in on-line approximate algorithms. Previous algorithms can compute approximate values of quantiles using constant or only weakly increasing memory footprint, but these previous algorithms cannot provide constant relative accuracy.  The new algorithm described here, the $t$-digest, provides constant memory bounds and constant relative accuracy while operating in a strictly on-line fashion.

\subsection{Previous work}

One early algorithm for computing on-line quantiles is described by Chen, et alia in \citep{Chen2000}.  In that work specific quantiles were computed by incrementing or decrementing an estimate by a value proportional to the simultaneously estimated probability density at the desired quantile.  This method is plagued by a circularity in that estimating density is only possible by estimating yet more quantiles.  Moreover, this work did not allow the computation of hybrid quantities such as trimmed means.

Munro and Paterson\citep{munro1980} provided an alternative algorithm to get a precise estimate of the median.  This is done by keeping $s$ samples from the $N$ samples seen so far where $s << N$ by the time the entire data set has been seen.  If the data are presented in random order and if $s = \theta(N^{1/2} \log N)$, then Munro and Paterson's algorithm has a high probability of being able to retain a set of samples that contains the median.  This algorithm can be adapted to find a number of pre-specified quantiles at the same time at proportional cost in memory.  The memory consumption of Munro-Paterson algorithm is, however, excessive if precise results are desired.  Approximate results can be had with less memory, however.  

A more subtle problem is that the implementation of Munro and Paterson's algorithm in Sawzall\citep{sawzall} and the Datafu library\citep{datafu} uses a number of buckets computed from the GCD of the desired quantiles.  This means that if you want to compute the $99$-th, $99.9$-th and $99.99$-th percentiles, ten thousand buckets are required, each of which requires the retention of many samples. We will refer to this implementation of Munro and Paterson's algorithm as MP01 in results presented here.

One of the most important results of the work by Munro and Paterson was a proof that computing any particular quantile exactly in $p$ passes through the data requires $\Omega(N^{1/p})$ memory. For the on-line case, $p=1$, which implies that on-line algorithms cannot guarantee to produce the precise value of any particular quantile. This result together with the importance of the on-line case drove subsequent work to focus on algorithms to produce approximate values of quantiles.

Greenwald and Khanna\citep{Greenwald-space-efficient-online-quantiles} provided just such an approximation algorithm that is able to provide estimates of quantiles with controllable accuracy. This algorithm (which we shall refer to as GK01 subsequently in this paper) requires less memory than Munro and Paterson's algorithm and provides approximate values for pre-specified quantiles.

An alternative approach is described by Shrivastava and others in \citep{qdigest}.  In this work, incoming values are assumed to be integers of fixed size. Such integers can trivially be arranged in a perfectly balanced binary tree where the leaves correspond to the integers and the interior nodes correspond to bit-wise prefixes. This tree forms the basis of the data structure known as a Q-digest.  The idea behind a Q-digest is that in the uncompressed case, counts for various values are assigned to leaves of the tree.  To compress this tree, sub-trees are collapsed and counts from the leaves are aggregated into a single node representing the sub-tree such that the maximum count for any collapsed sub-tree is less than a threshold that is a small fraction of the total number of integers seen so far.  Any quantile can be computed by traversing the tree in left prefix order, adding up counts until the desired fraction of the total is reached.  At that point, the count for the last sub-tree traversed can be used to interpolate to the desired quantile within a small and controllable error.  The error is bounded because the count for each collapsed sub-tree is bounded.

The salient virtues of the Q-digest are
\begin{itemize}[nosep, topsep=-10pt]
\item the space required is bounded proportional to a compression factor $k$
\item the maximum error of any quantile estimate is proportional to $1/k$ and
\item the desired quantiles do not have to be specified in advance.
\vspace{10pt}
\end{itemize}

On the other hand, two problems with the Q-digest are that it depends on the set of possible values being known in advance and produces quantile estimates with constant error in $q$. In practice, this limits application of the Q-digest to samples which can be identified with the integers. Adapting the Q-digest to use a balanced tree over arbitrary elements of an ordered set is difficult.  This difficulty arises because rebalancing the tree involves sub-tree rotations and these rotations may require reapportionment of previously collapsed counts in complex ways.  This reapportionment could have substantial effects on the accuracy of the algorithm and in any case make the implementation much more complex because the concerns of counting cannot be separated from the concerns of maintaining a balanced tree.  
\subsection{New contributions}
The work described here introduces a new data structure known as the $t$-digest which is formed by clustering real-valued samples. The $t$-digest differs from more well-known forms of clustering such as $k$-means in that the samples are from $\mathbb R^1$ rather than taken from any space with a metric and because the clusters are limited by size in a special way. The $t$-digest differs from previous structures designed for computing approximate quantiles in several important respects. First, although data is clustered and summarized in the $t$-digest, the range of data included in different clusters may overlap. Second, the bins are summarized by a centroid value and an accumulated weight representing the number of samples contributing to a bin rather than by the upper and lower bounds of the bin. Third, the samples are accumulated in such a way that only a few samples contribute to bins corresponding to extreme quantiles so that relative error is bounded instead of maintaining constant absolute error as with previous methods.

With the $t$-digest, accuracy for estimating the $q$ quantile is nearly constant relative to $q(1-q)$.  This is in contrast to earlier algorithms which had errors independent of $q$.  The relative error bound of the $t$-digest is convenient when computing quantiles for $q$ near $0$ or $1$ as is commonly required.  As with the Q-digest algorithm, the accuracy/size trade-off for the $t$-digest can be controlled by setting a single compression parameter $\delta$ with the amount of memory required proportional only to $\Theta(\delta)$. 

\section{The $t$-digest}
A $t$-digest is generated by clustering real-valued samples and retaining the mean and number of samples for each cluster. This clustering can then be used to estimate quantile-related statistics with particularly high accuracy near the tails of a distribution. Algorithmically, there are two important ways to form a $t$-digest from a set of numbers. One version keeps a buffer of incoming samples. When the buffer fills, the contents are sorted and merged with the centroids computed from previous samples. This merging form of the $t$-digest algorithm has the virtue of allowing all memory structures to be allocated statically. On an amortized basis, this buffer-and-merge algorithm can be very fast especially if the input buffer is large. The other major $t$-digest algorithm is more akin to traditional clustering algorithms where new samples are added one at a time to whichever cluster is nearest.  Both algorithms are described here and implementations for reference implementations of both are available as open source software. 

\subsection{The basic concept}
Take a $n$ samples $\lbrace x_1 \ldots x_n \rbrace = X \subset \mathbb R^1$. Define a digest  as a partition of $X$ consisting of disjoint  sets of samples $\pi_i \subset X$. The subsets $\pi_i$  in this partition are referred to as clusters. The number of samples in a cluster $\mathcal C$ is written as $|\mathcal C|$ and is typically referred to as the weight of the cluster.
 
Each cluster $\mathcal C$ in a digest has $|C|>0$ samples and an associated mean $\overline {\mathcal C} = \sum_{x\in \mathcal C} x / |\mathcal C|$. The clusters in the digest can be partially ordered according to their means. Clusters with identical means are assigned an arbitrary fixed ordering so that we can index the clusters consistently. Given this complete ordering, we define a left and right weight for each cluster $\mathcal C_i$ as the sum of the weights of clusters to the left and right of $\mathcal C_i$. That is
\begin{align*}
\mathcal W_{\mathrm{left}}(\mathcal C_i) = \sum_{j < i} |C_j| \\
\mathcal W_{\mathrm{right}}(\mathcal C_i) = \sum_{j > i} |C_j|
\end{align*}

Note that so far we have no constraint about the ordering of the elements in different point sets. We refer to a digest as strongly ordered if 
\begin{equation}\label{eq:weak-ordering}
 i > j \implies x \ge y \mathrm{\,for\,} x \in \pi_i \mathrm{\,and\,} y \in \pi_j
 \end{equation}
. We refer to a digest as weakly ordered if
\begin{equation}\label{eq:strong-ordering}
 i > j+\Delta \implies x \ge y \mathrm{\,for\,} x \in \pi_i \mathrm{\,and\,} y \in \pi_j
 \end{equation}
for some positive offset $\Delta\ge 1$.

Such a digest is a $t$-digest if every cluster has unit weight, or has weight bounded using a scale function as defined below. A $t$-digest is called fully merged if no two consecutive clusters can be combined without violating the weight bound.

\subsection{A simple start}

Suppose that all of the samples $X=x_1 \ldots x_n$ are presented in ascending order with ties broken arbitrarily. Since the samples are ordered, we can use the index of each sample to determine the value of the empirical quantile for any new value. 

If we form a trivial digest where each cluster has a single point, this digest will be a strongly ordered $t$-digest, but will likely not be fully merged.

We can group consecutive samples greedily from left to right into sub-sequences of consecutive samples, $X = \lbrace \pi_1 | \pi_2 | \ldots | \pi_m \rbrace$ where $\pi_i = \lbrace x_{\mathtt {left}(i)} \ldots x_{\mathtt{right}(i)} \rbrace$ such that each $\pi_i$ has as many samples as possible subject to the size bound. The result is a $t$-digest and is also be a strongly ordered and fully merged.

One key idea of the $t$-digest is that the size of each cluster is chosen so that it is small enough to get accurate quantile estimates by interpolation, but large enough so that we don't wind up with too many clusters. Importantly, we force clusters near both ends to be small while allowing sub-sequences in middle to be larger in order to get fairly constant relative accuracy. The other key idea is that there are practical algorithms that allow us to build $t$-digests incrementally, and although the result may not quite be strongly ordered, it will be close enough to be very useful.

\subsection{The size bound for clusters}
The size bound for $t$-digests is imposed using a scale function that forces clusters near the beginning or end of the digest to be small, possibly containing only a single sample. The scale function is chosen to provide an appropriate trade-off between very accurate quantile estimate in the tails of a distribution, reasonable accuracy near the median while keeping the number of clusters as small as possible.

\begin{figure}[htbp] 
   \centering
   \includegraphics[width=2.5in]{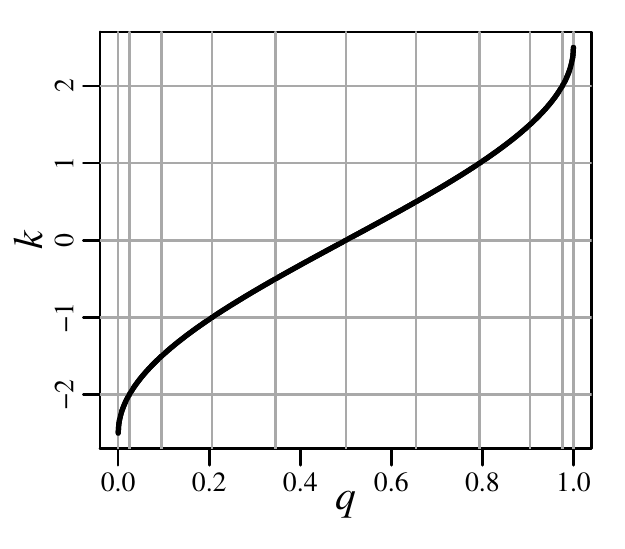} 
   \caption{The scale function translates the quantile $q$ to the scale factor $k$ in order to give variable size steps in $q$. Limiting cluster sizes allows better accuracy near $q=0$ or $q=1$. }
   \label{fig:k-q-plot}
\end{figure}

To limit cluster size in this way, we define the scale function as a non-decreasing function from quantile $q$ to a notional index $k$ with compression parameter $\delta$. A one commonly used  scale function for the $t$-digest is 
\begin{equation}
k_1(q) = \frac \delta {2\pi}  { {\sin^{-1} (2q-1)} }    
\end{equation}
As with any scale function, $k_1$ is non-decreasing. It has minimum value  $k(0)=-\delta/4$ and maximum value  $k(1)=\delta/4$.  Figure \ref{fig:k-q-plot} shows the relationship between $k$ and $q$ for $\delta=10$. 
In this figure, the horizontal lines are spaced uniformly at integer values of $k$. Vertical lines are drawn from the intersection of the scale function with these evenly spaced horizontals. Note how much closer together these vertical lines are near $q=0$ and $q=1$.

The scale function provides the necessary mechanism to define the size bound of a $t$-digest. Every cluster $\mathcal C$ with more than one sample should have the $k$-size (written as $|\mathcal C|_k$) at most $1$,
\begin{equation}
\label{eq:bounded-size}
|\mathcal C|_k = k \left (q_{\mathrm {right}} \right) - k \left ( q_{\mathrm {left}} \right) \le 1
\end{equation}
where
\begin{align*}
q_{\mathrm {left}} &= {\mathcal W}_{\mathrm{left}}(\mathcal C)/n \\
q_{\mathrm {right}} &= q_{\mathrm {left}} + {| \mathcal C | / n} 
\end{align*}
In a fully merged $t$-digest adjacent clusters, cannot be merged because the result would be too big 
\begin{equation}
\label{eq:fully-merged}
|\mathcal C_i \cup \mathcal C_{i+1} |_k = |\mathcal C_i|_k + |\mathcal C_{i+1}|_k > 1
\end{equation}
Together, the conditions expressed in equations \ref{eq:bounded-size} and \ref{eq:fully-merged} imply that for a fully merged $t$-digest using $k_1$ with $n \gg \delta$ samples, the number of clusters $m$ will be at most $ \left \lceil\delta \,\right \rceil $.  The lower bound  on the number of clusters is near $ \lfloor\delta/2\rfloor $.

\subsection{The effect of a scale function}

The only really necessary characteristic of a scale function up to now is that be non-decreasing. As such,  a linear scale function $k_0(q) = \delta q/2$ can be used instead of  $k_1$. Using a linear scale function results in nearly uniform cluster sizes and constant absolute error in $q$  which is similar to the behavior of most previously reported quantile approximation algorithms. 

\begin{figure}[htb] 
   \centering
   \includegraphics[height=2.in, clip]{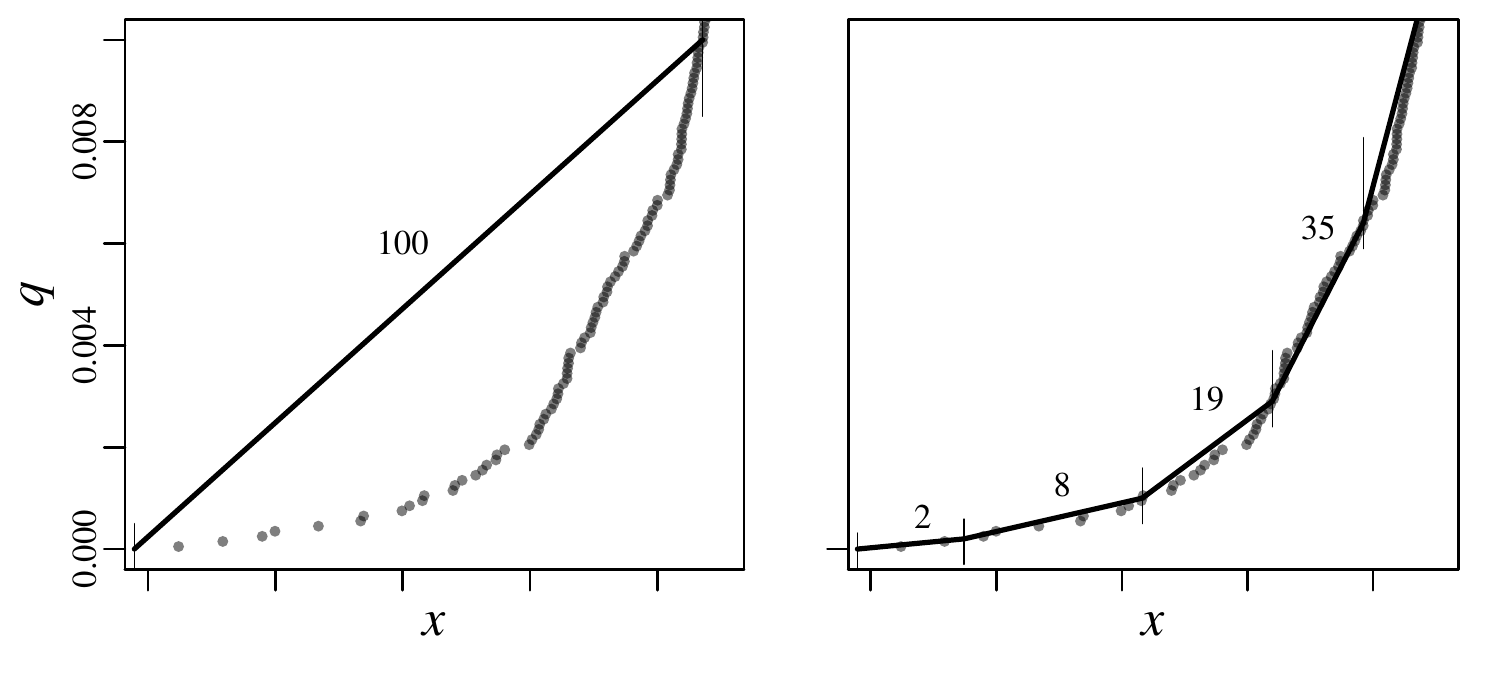} 
   \caption{The left panel shows linear interpolation of the cumulative distribution function near $q=0$ with $100$ equal sized bins applied to $10,000$ data points sampled from an exponential distribution. The right panel shows the same interpolation with variable size bins as given by a strongly ordered $t$-digest with $\delta=100$. The numbers above the data points represent the number of points in each bin. }
   \label{fig:linear-interpolation}
\end{figure}

Figure \ref{fig:linear-interpolation} shows how  estimating $q$ near tails is improved for a strongly ordered, fully merged $t$-digest constructed with a scale function that keeps clusters small near extreme values of $q$. 
This figure shows roughly the first percentile of 10,000 data points sampled using $x \sim \log u$ where $u \sim \mathrm{Uniform}(0,1)$. In the left panel, the data points have been divided into 100 clusters, each with 100 data points, of which only the left-most bin is visible. The use of equal sized bins means that the interpolated value of $q$ in the left panel of the figure has a substantial and unavoidable error. The right panel, on the other hand, shows a $t$-digest with roughly the same number of bins ($102$ instead of $100$), but with many fewer points in bins near  $q=0$. 

Of course, since every sample must be in some bin, filling some bins with fewer than $100$ samples requires that other bins must have more than $100$ samples or we have to have more bins. Note, however, that having fewer samples in the bins near $q=0$ or $q=1$ improves accuracy substantially, while increasing the bin sizes near $q=1/2$ degrades accuracy only modestly. In this particular example, the first bin has only 2 samples and thus zero error and the second bin has only 10 samples giving much smaller error than a cluster of $100$ samples would give. The clusters near $q=1/2$, on the other hand, have about $1.6$ times more samples than the uniform case, increasing errors by only a modest factor relative to equal sized bins. The overall effect is that quantile estimation accuracy is dramatically improved at the extremes but only modestly impaired near the median. Exactly how much accuracy at the tails is improved depends crucially on how close the $t$-digest is to being strictly ordered.

\subsection{Merging independent $t$-digests}
In the previous section, the algorithm for forming a $t$-digest took a set of samples as inputs. Nothing, however, would prevent the algorithm from being applied to a set of weighted samples. As long as individual weights are smaller than the size limit imposed by the scale function, the result can still be a well-formed $t$-digest.

If we form independent $t$-digests $t_X$ and $t_Y$ from separate sequences $X$ and $Y$, these $t$-digests can clearly be used to estimate quantiles of $X \cup Y$ by separately computing quantiles for $X$ and $Y$ and combining the results. It is much more useful, however, if we choose the scale factor so that it guarantees that the ordered union of two $t$-digests will still be a valid $t$-digest. The scale factors $k_0$ and $k_1$ both have this property as do the $k_2$ and $k_3$ scale functions described later, but there is no guarantee that a candidate scale function will have this property.

We can force a  digest formed by merging other digests to be fully merged by combining  centroids wherever consecutive clusters taken together meet the size bound. The resulting $t$-digest will not necessarily be the same as if we had computed a $t$-digest $t_{X \cup Y}$ from all of the original data at once even though it will meet the same size constraint. In particular, even if $t_X$ and $t_Y$ are individually strongly ordered, their union $t_X \cup t_Y$ will, in general, only be weakly ordered. This can happen, for instance, when there are centroids in very nearly the same position in the two digests being merged. This loss of strictly ordering makes it difficult to compute rigorous error bounds.

In practice, $t$-digests formed by merging other digests still do produce accurate quantile estimates, even with highly structured data sets such as ordered data or data with large numbers of repeated values, but we do not yet have a good understanding how weak the ordering is in merged digests. The section on Empirical Results examines how these issues play.

The ability to merge $t$-digests makes parallel processing of large data-sets relatively simple since independent $t$-digests can be formed from disjoint partitions of the input data and then combined to get a $t$-digest representing the complete data-set.

A similar divide and conquer strategy can be used to allow $t$-digests to be used in OLAP systems. The idea is that queries involving quantiles of subsets of data can be approximated quickly by breaking the input data set into subsets corresponding to each unique combination of selection factors. A single $t$-digest is then pre-computed for each unique combination of selection factors. To the extent that all interesting queries can be satisfied by disjoint unions of such primitive subsets, the corresponding $t$-digests can be combined to compute the desired result.

\subsection{Progressive merging algorithm}
The observation that merging $t$-digests gives good accuracy suggests a practical algorithm for constructing a $t$-digest from a large amount of data. The basic idea is to collect data in a buffer. When the buffer fills, or when a final result is required, sort the buffer contents together with any previously created centroids, and make a single pass to merge points or centroids  whenever the size limits are satisfied by the combination. With an arbitrarily large buffer, this algorithm reduces to the original approach for constructing a $t$-digest from sorted data since the overall effect is simply a single merging pass through all the data. For smaller buffer sizes, however, many merge passes are required to process a large amount of data. 

The algorithm for merging a buffer's worth of samples into an existing set of centroids is shown as Algorithm \ref{alg:merge}. Note that the check on the size bound is optimized to only require only as many evaluations of $k$ and $k^{-1}$ during the merge as there are output values in $C'$. This is done by computing the bounding value $q_{\mathrm {limit}} = k^{-1}(k(q) + 1)$ each time a new centroid is emitted. This allows the conditional to be couched in terms of comparisons of $q$ so that if many points are merged, no additional calls to $k$ are needed. If $n \gg  \left \lceil \delta \right \rceil$, this can result in a considerable speedup because evaluating a scale function or its inverse typically involves calls to $\log$ or $\sin^{-1}$. Note also that this algorithm only uses four statically allocated arrays of primitive values, avoiding all dynamic allocation and structure boxing/unboxing.
\begin{algorithm}[ht]
\SetKw{KwTo}{in}\SetKwFor{For}{for}{\string:}{}
\SetKwIF{If}{ElseIf}{Else}{if}{:}{elif}{else:}{}
\SetKwFor{While}{while}{:}{fintq}
\SetKwFor{For}{for}{\string:}{}
\SetNoFillComment
\KwIn{Sequence $C = [c_1 \ldots c_m]$ a $t$-digest containing real-valued, weighted centroids with components $\mathtt{sum}$ and $\mathtt{count}$ arranged in ascending order by mean, data buffer  $X = { x_1,\ldots x_n}$ of real-valued, weighted points, and compression factor $\delta$}
\KwOut{New ordered set $C'$ of weighted centroids forming a $t$-digest} 
$X \gets \mathtt{sort}(C \cup X)$\;
$ S = \sum_i x_i.\mathtt{count}$\;
$C' = \lbrack \, \rbrack, q_0 = 0$\;
$q_{limit}=k^{-1}(k(q_0, \delta)+1, \delta)$\;
$\sigma = x_1$\;
\For{$i \in 2\ldots(m+n)$} {
  $q = q_0 + (\sigma.\mathtt{count} + x_i.\mathtt{count})/S$\;
  \If {$q \le q_{limit}$} {
      $\sigma \gets \sigma + x_i$\;
  } \Else {
      $C'\mathtt{.append}(\sigma)$\;
      $q_0 \gets q_0 + \sigma.\mathtt{count}/S$\;
      $q_{limit} \gets k^{-1}(k(q_0, \delta)+1, \delta)$\;
      $\sigma \gets x_i$\;
  }
} 
$C'\mathtt{.append}(\sigma)$\;
\Return $ C' $\\
\caption{Merging new data into a $t$-digest\label{alg:merge}}
\end{algorithm}

The run-time cost of the merge variant of the $t$-digest is a mixture of the frequent buffer inserts  and the rare  merges. The buffer inserts are very fast since they consist of an array write, index increment and an overflow test. The merges consist of the buffer sort and the merge itself. The merge involves a scan through both the buffer and the existing centroids plus a number of calls to the scale function roughly equal to the size of the result which is bounded by $ \lceil \delta \rceil$ for all common scale functions. If $c_1$ is the input buffer size, the dominant costs are the sort and the scale function calls so the amortized cost per input value is roughly $C_1 \log c_1 + C_2 \lceil \delta \rceil / c_1$ where $C_1$ and $C_2$ are parameters representing the sort and scale function costs respectively. This overall amortized cost has a minimum where $c_1 \approx \delta\, C_2  / C_1$. The constant of proportionality should be determined by experiment, but micro-benchmarks indicate that $C_2 / C_1$ is in the range from $5$ to $20$ for a single core of an Intel i7 processor. In these micro-benchmarks, increasing the buffer size to $10 \lceil \delta \rceil$ dramatically improves the average speed but further buffer size increases have much less effect. 

Further optimization in the case of the scale function $k_1$ is possible by speeding up the evaluation of $\sin^{-1}$. It is difficult to build a very fast and still high quality approximation of $\sin^{-1}$ over the entire domain $[0,1]$, but by limiting the domain where the approximation is used to $q \in [\epsilon, 1-\epsilon]$, where $\epsilon \approx 0.01$, simple and fast approximations are available.

It is also possible to avoid evaluation of $k$ and $k^{-1}$ by estimating the maximum number of samples that can be in each candidate cluster directly from $q$. Such estimates typically under-estimate the number of samples allowed, especially near the tails, but the size of the $t$-digest is not substantially increased and accuracy can be somewhat increased.

Finally, in practice, merge-based algorithms as shown here can be subject to problems due to the fact that samples are always considered in ascending order as they are merged with existing centroids. This can cause centroids near $q=0$ to grow in an asymmetric fashion with the effect that the centroid drifts a bit as the digest is constructed. This drift means that the resulting digest is more weakly ordered than desired. This problem can be mitigated to a degree by alternating the scan order in the merge (ascending in one pass, descending the next). 

Another approach is to use a stratified merge in which a temporarily larger value of the compression parameter $\delta$ is used while points are being added. When the digest is stored, a final merge pass is made to fully consolidate the digest with the final compression parameter. To some degree, this can be framed in the context of a three-way tradeoff between space, speed and accuracy. For highest speed with constant memory use, a large buffer and a smaller value of $\delta$ can be used. For highest accuracy, a large buffer and stratified merge can be used. For smallest memory, a smaller buffer and a single value of $\delta$ should be used. 

To give a specific example of this trade-off, for high accuracy, a buffer size of $10\times\delta$ would be used as is commonly done but merging could be done with a compression parameter of $3\times\delta$. This would require no more memory than typically used, but would cause merging to be done about 20\% more often, and would result in 3 times as many centroids being retained between merges which would make each merge a bit more expensive. If the resulting digest is only weakly ordered, say with ordering parameter $\Delta \le 3$ due to being constructed incrementally, the final merge is likely to reduce the digest nearly to a strongly ordered state as the number of centroids and ordering parameter are both reduced by roughly a factor of 3 to 1. In practice, the reduction in ordering parameter results in dramatically smaller errors.

Stratified merging is also useful when $t$-digest is used in parallel implementations. This is described in more detail in the section on parallel computing.

\subsection{The clustering variant}
If we allow the buffer in the merging variant of the $t$-digest algorithm to contain just a single element so that merges take place every time a new point is added, the algorithm takes on a new character and becomes much more like clustering than buffering and merging.

The basic outline of the clustering algorithm for constructing a $t$-digest is quite simple.  An initially empty ordered list of centroids, $C = [ c_1 \ldots c_m ]$ is kept.  Each centroid consists of a mean and a count.  To add a new value $x_n$ with a weight $w_n$, the set of centroids is found that have minimum distance to $x_n$.  This set is reduced by retaining only centroids whose $k$-size after adding $w_n$ would meet the size bound.  If more than one centroid remains, the one with maximum weight is selected.  If an acceptable centroid is found, then the new point, $(x_n,w_n)$, is added to that centroid. If no satisfactory centroid is found then $(x_n,w_n)$ is used to form a new centroid with weight $w_n$.

This clustering variant is shown as Algorithm \ref{alg:full}.
 \begin{algorithm}[htb]
\SetKw{KwTo}{in}\SetKwFor{For}{for}{\string:}{}
\SetKwIF{If}{ElseIf}{Else}{if}{:}{elif}{else:}{}
\SetKwFor{While}{while}{:}{fintq}
\SetKwFor{For}{for}{\string:}{}
\SetNoFillComment
\KwIn{Ordered set of weighted centroids $C = \lbrace \rbrace$, sequence of real-valued, weighted points $X = \lbrace (x_1, w_1),\ldots (x_N, w_N)\rbrace$, accuracy tolerance $\delta$, and limit on excess growth $K$ typically in the range $3\ldots10$}
\KwOut{final set $C=[c_1 \ldots c_m]$ of weighted centroids} 
\For{$(x_n, w_n) \in X$} {
  $z = \min | c_i.\mathtt{mean} - x |$\;
  $S = \lbrace c_i  :  |c_i.\mathtt{mean} - x| = z \wedge |c_i+w_1|_k < 1 \rbrace $\;
  \If {$|S| > 0$} {
       \tcp{sort by descending distance}
       $S.\mathtt{sort}( key=\lambda (c) \lbrace -c.\mathtt{sum} \rbrace)$\;
       $c \gets S.\mathtt{first()}$ \;
       $c.\mathtt{count} \gets c.\mathtt{count} + w_n$\;
       $c.\mathtt{mean} \gets c.\mathtt{mean} + (x_n- c.\mathtt{mean})/ c.\mathtt{sum}$\;
     } \Else {
      $C \gets C + (x_n, w_n)$\;
      }
      \If {$|C| > K\delta$} {
         $C \gets \mathtt{merge}( C, \lbrace\rbrace ) $\;
       }
} 
$C \gets \mathtt{merge}( C, \lbrace\rbrace ) $\;
\Return $ C $\\
\caption{Construction of a $t$-Digest by clustering \label{alg:full}}
\end{algorithm}

Certain insertion orders can cause the number of centroids to increase without bound. For instance, if the values of $X$ are in ascending or descending order, $C$ will contain as many centroids as samples inserted.  This happen because each new value of $X$ is always a new maximum (or minimum) and thus will always form a new centroid.  To avoid this, the clusters can be consolidated using Algorithm \ref{alg:merge} whenever the number of clusters grows excessively.  

\subsection{Alternative scale functions}
The $k_1$ and $k_0$ scale functions that we have mentioned are not the only ones possible. In fact, there are two additional important functions that provide important accuracy/size trade-offs. Altogether, these four functions serve as useful scale functions
\begin{align}
k_0(q) &= \frac \delta 2 q \\
k_1(q) &= \frac \delta {2\pi}  \sin^{-1}(2q-1)   \\
k_2(q) &= \frac \delta {4 \log n/\delta + 24} \log {\frac q {1-q}} \\
k_3(q) &= \frac \delta {4\log n/\delta + 21}\begin{cases}
\quad \log 2q & \text{if  } q \le 1/2 \\
- \log 2(1-q) & \text{if  } q > 1/2
\end{cases}
\end{align}

Each of these scale functions to build a $t$-digest gives the following two guarantees
\begin{itemize}
\item
The number of centroids in a fully merged $t$-digest is bounded by $\ceil\delta$ as the number of samples increases without bound (for $k_2$ and $k_3$ as given above, this only applies up to $10^{60}$ samples when the number of clusters starts growing roughly with $\log \log n$)
\item
The union of the centroids from multiple $t$-digests will meet the size bound and thus be a $t$-digest for a value of $\delta$ equal to the smallest $\delta$ of any of the individual digests.
\end{itemize}

It is also possible to use unnormalized versions of $k_2$ and $k_3$:
\begin{align}
k'_2(q) &=  \delta  \log {\frac q {1-q}} \\
k'_3(q) &=  \delta \begin{cases}
\quad \log 2q & \text{if  } q \le 1/2 \\
- \log 2(1-q) & \text{if  } q > 1/2
\end{cases}
\end{align}
With these denormalized scale functions the number of centroids in a $t$-digest will grow roughly with $\log n$.

The major benefit of $k_2$ and $k_3$ scale functions is that they constrain the size of clusters near $q=0$ and $q=1$ even more strenuously than $k_1$ does. Whereas, clusters are limited in size proportionally to $\sqrt {q (1-q)}$ for $k_1$, $k_2$ and $k_3$ limit cluster sizes proportional to $q(1-q)$ and $\min(q,1-q)$ respectively where $q$.

This more rigorous regulation of cluster size means that errors in interpolation due to weak ordering of the clusters in a digest or due to non-uniform distribution of samples for clusters will be compensated by having smaller clusters near extreme values of $q$. The $k_3$ scale function, in particular, will drive errors to zero near the extremes by forcing more clusters near $q=0$ or $q=1$ to have only a single sample.

\subsection{Interpolation of the cumulative distribution function}
The information in a fully-merged $t$-digest is not sufficient, in general, to precisely determine the empirical quantile $q$ of any particular value $x$ because information is lost as samples are clustered together. The only exception to this is in the case of clusters containing only a single sample where the centroid is the same as the single sample.

The approach taken in current implementations of the $t$-digest is to make some simplifying assumptions about the distribution of samples in the original data. These are:
\begin{enumerate}
\item The samples for a cluster are evenly split into samples that are to the left of the centroid and those to the right if there is more than one sample. 
\item The samples between two clusters are uniformly distributed between them. 
\end{enumerate}
The first assumption is obviously the best we can do with no further information about the expected distribution. The second assumption is true asymptotically for continuous distributions as the number of clusters in a $t$-digest increases if the samples associated with each cluster do not extend beyond the neighboring centroids. In practice, clusters in a $t$-digest are sufficiently localized and quantile accuracy is good even for highly skewed distributions.

Based on these assumptions, we have four important cases. These are a) interpolation between clusters each with more than one sample, b) interpolation between a multi-sample cluster and a cluster with just a single sample, c) interpolation between clusters each with just a single sample and d) interpolation for the first and last cluster.

\begin{figure}[htb] 
   \centering
   \includegraphics[width=3in]{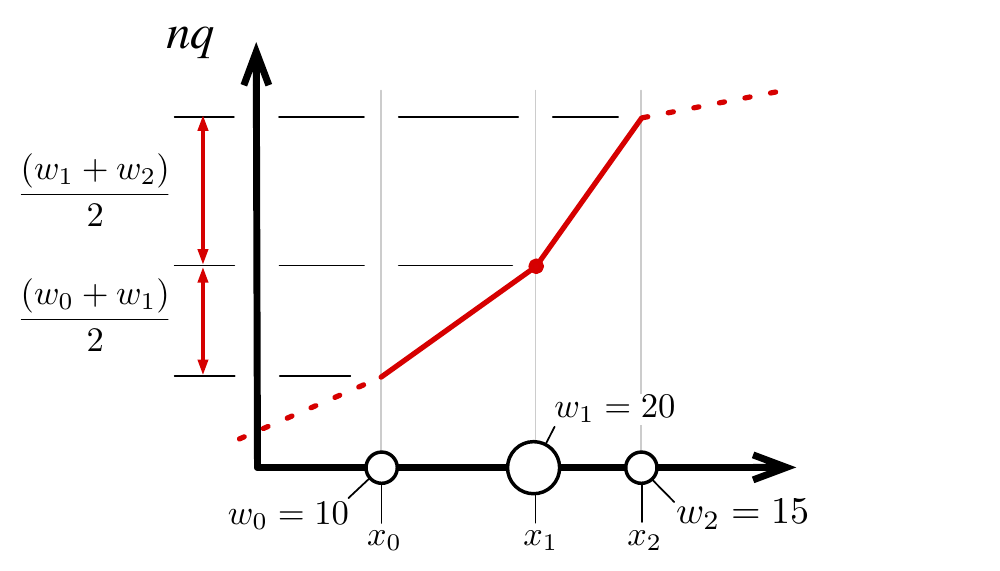} 
   \caption{Interpolation of the empirical cumulative distribution function between centroids of clusters with more than one sample is done by assuming half of the points for each centroid are to the left of the centroid and half are to the right. }
   \label{fig:interpolation}
\end{figure}

Between two clusters where both have more than one sample, the empirical cumulative distribution function is approximated using a linear approximation between two consecutive centroids that allocates half the weight of each centroid to the interval between them. This is equivalent to saying that the the mid-point of the cluster is at the centroid. This is illustrated in Figure \ref{fig:interpolation}. 

\begin{figure}[tb] 
   \centering
   \includegraphics[width=3in]{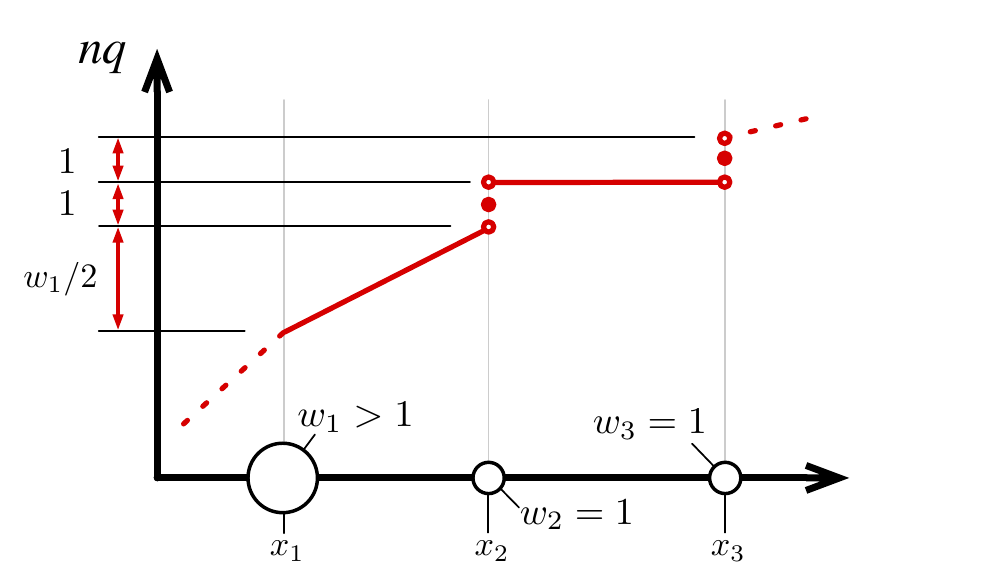} 
   \caption{Clusters with a single sample are handled specially. Adjacent to a normal cluster, as is the case between $x_1$ and $x_2$, interpolation is done assuming $w_1/2$ samples occur between $x_1$ and $x_2$, and the single sample at $x_2$. That single sample causes the cumulative distribution to step to the mid-point of the individual sample at $x_2$. Between singleton clusters $x_2$ and $x_3$, the cumulative distribution is given a constant value until it steps again at $x_3$.}
   \label{fig:singletons}
\end{figure}

When clusters have a single sample, however, this interpolation can be improved by noting that this single sample is located exactly at the centroid for that cluster. We can use that to improve our interpolation as illustrated in Figure \ref{fig:singletons}. 
When the neighboring cluster has multiple samples, we can interpolate almost as before, but when the next cluster is itself a singleton, we can avoid interpolation entirely. 

\begin{figure}[htb] 
   \centering
   \includegraphics[width=3in]{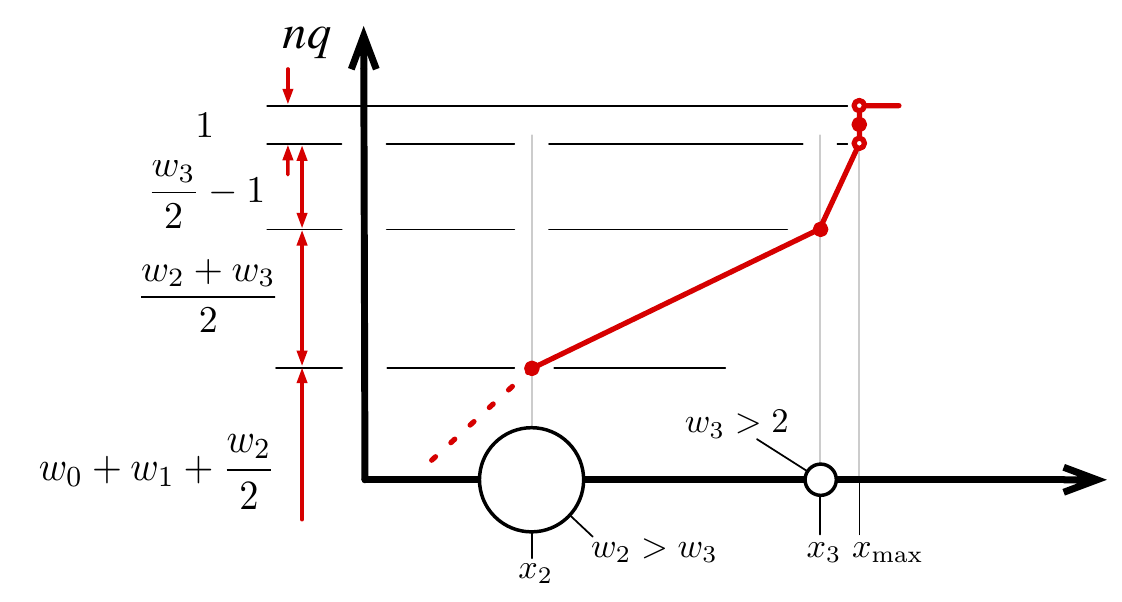} 
   \caption{For the last clusters, if the cluster weight is greater than $2$, we can use the fact that a singleton must occur at $x_{\mathrm {min}}$ or $x_{\mathrm {max}}$ to improve interpolation accuracy. The first cluster is handled by reversing these steps.}
   \label{fig:endpoint}
\end{figure}

The final case to be considered is that of the first or last cluster in the digest. 
When such a terminal cluster has only one sample, then we know that $x_{\mathrm {max}}$ is due exactly to that single sample. Where the cluster has two samples, we know that those two samples were at $x_{\mathrm {max}}$ and $x_n-(x_{\mathrm {max}}-x)$ and thus they can be treated as two singletons. Where there are more than two samples, however, it is advantageous to treat the cluster as having half the samples before the centroid and half after. For the half of the samples that come after, one of those samples must be at $x_{\mathrm {max}}$ and thus it can be treated as a singleton with the remainder being interpolated. All of this discussion is reversed, of course, first for last and left for right when treating the first cluster. This is shown in Figure \ref{fig:endpoint} for the last cluster in a digest.

For scale functions $k_2$ and $k_3$ the first and last clusters will always be singletons and thus will never need this special handling. For the $k_1$ scale function, however, the first and last clusters may well have more than one sample thus making it desirable to interpolate out beyond the centroid. 

\section{Empirical Assessments}
\begin{figure}[htb] 
   \includegraphics[width=5.5in]{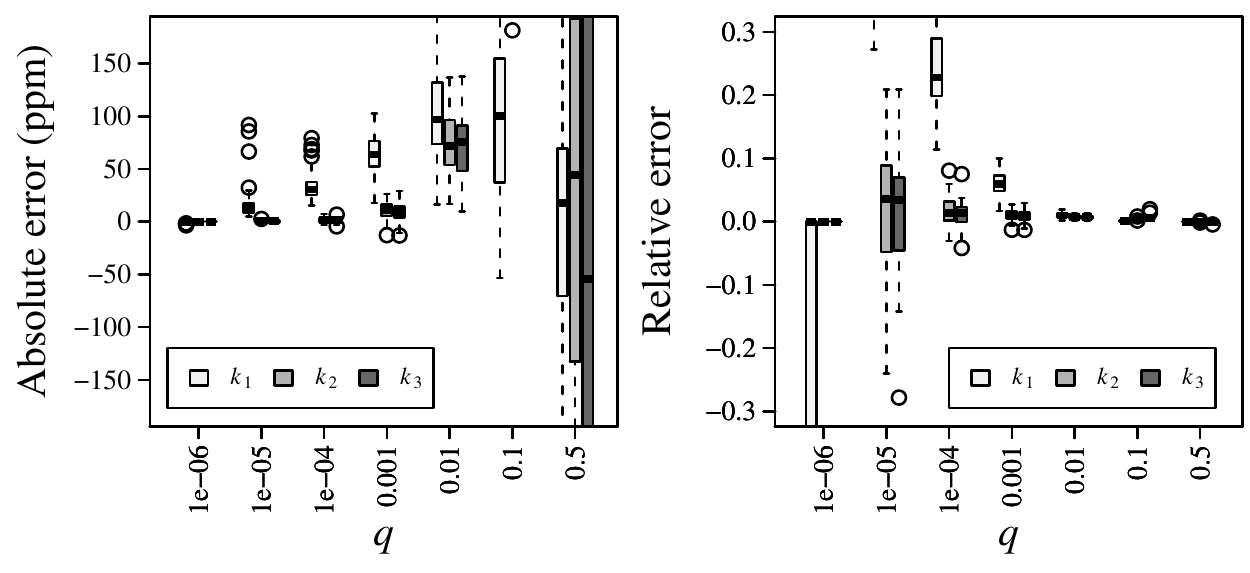} 
   \caption{The left panel shows absolute error in parts per million of estimations of quantile with $\delta = 100$ for scale functions $k_1$, $k_2$ and $k_3$. Errors are computed by taking $10^6$ samples from a uniform distribution and comparing estimates from a $t$-digest against exact quantiles computed from the original samples. This was repeated 50 times to get a sense of variability. The left panel shows absolute error in parts per million and the right panel shows relative error.}
   \label{fig:by-scale}
\end{figure}

Figure \ref{fig:by-scale} examines the accuracy of quantile estimation with a $t$-digest.  As intended, absolute error in quantile estimation is small and decreases near $q=0$ and $q=1$ (left panel) for scale functions $k_1$, $k_2$ and $k_3$. Though it does not show well in this figure because of the vertical scale, in the mid-range where $q \in [0.1, 0.9]$, $k_1$ provides somewhat better absolute error than $k_2$ and $k_3$. 

All scale functions show a pattern of decreasing absolute error for more and more extreme values of $q$, but $k_2$ and $k_3$ achieve single digit part per million errors for $q\le 0.001$ or $q\ge0.999$ which is surprisingly good since only about 50 to 60 centroids are retained in total for $\delta=100$. 

In spite of the decrease in the absolute error with all three scale functions as $q$ becomes more extreme, the relative error is not as well controlled. This is shown in right panel of Figure \ref{fig:by-scale}. Scale functions $k_2$ and $k_3$ exhibit good control of relative error down to $q=0.001$ but for $q= 10^{-5}$ and $q=10^{-4}$, relative error increases a bit. For  smaller values of $q$, all relevant centroids have only a single sample and thus errors go to zero. For $k_1$, the situation is not as favorable; relative error increases dramatically starting at $q=0.001$ and is off scale by $q=10^{-5}$.
\subsection{Degree of overlap of $t$-digest clusters}
\begin{figure}[htb] 
   \includegraphics[width=4.5in]{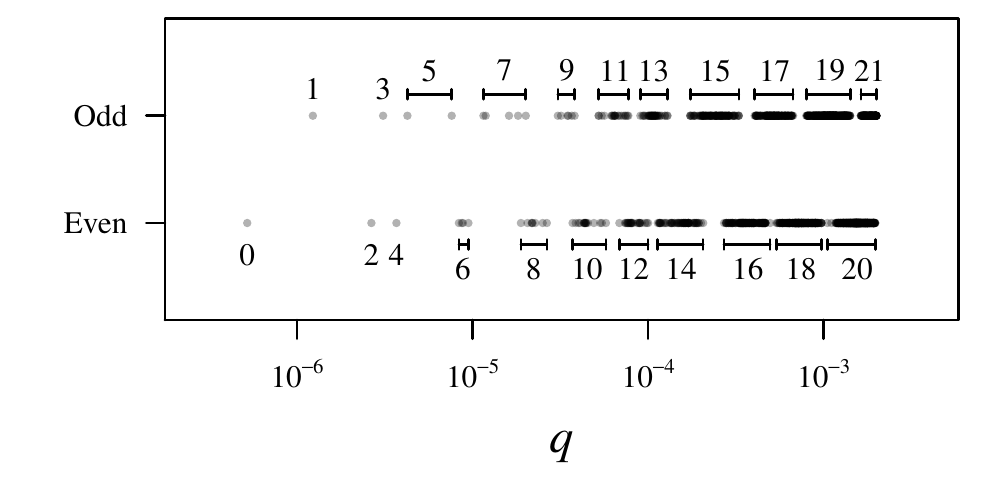} 
   \caption{The first few clusters in a typical $t$-digest show that while the digest is not strictly ordered, only adjacent clusters overlap. This plot shows roughly the first 20 clusters of the digest with even-numbered clusters below and odd-numbered ones above. Note how odd clusters overlap with adjacent even clusters, but never with adjacent odd clusters. This corresponds to weak ordering with $\Delta=1$  }
   \label{fig:cluster-spread}
\end{figure}

Figure \ref{fig:cluster-spread} shows how effective the stratified merging strategy  is at producing high quality clusters. This requires about 20\% more merge operations and  each merge operation is typically less than twice as expensive. The resulting clusters are not quite strictly ordered, but overlap between clusters is limited to adjacent clusters.
The digest shown here was produced by inserting 
a total of $10^6$ samples using a working compression of $\delta = 316$ and alternating the direction of merges. The final digest was produced by doing a final merge step with $\delta = 100$ resulting in a roughly $3:1$ merge of the working clusters. With unidirectional merge steps and keeping $\delta=100$ during construction, the overlap between clusters commonly extends to the adjacent 3-4 clusters, substantially degrading accuracy.

\subsection{Persisting $t$-digests}
For the accuracy setting and test data used in these experiments, the $t$-digest contained $50-60$ centroids.  The results of $t$-digest can thus be stored by storing this many centroid means and weights.  If centroids are kept as double precision floating point numbers and counts kept as 4-byte integers, the $t$-digest resulting from from the accuracy tests described here would require less than 800 bytes of storage.

This size can be decreased, however.  One simple option is to store differences between centroid means and to use a variable byte encoding to store each cluster size.  The differences between successive means are are about two orders of magnitude smaller than the means themselves so using single precision floating point to store these differences can allow the $t$-digest from the tests described here to be stored in under 500 bytes while still regaining nearly 10 significant figures of accuracy in the means.  This is roughly equivalent to the maximum precision possible with a Q-digest operating on 32 bit integers, but the dynamic range of $t$-digests will be considerably higher and the accuracy is considerably better.

\subsection{Space/Accuracy Trade-off}
\begin{figure}[p] 
   \centering
   \includegraphics[width=4in]{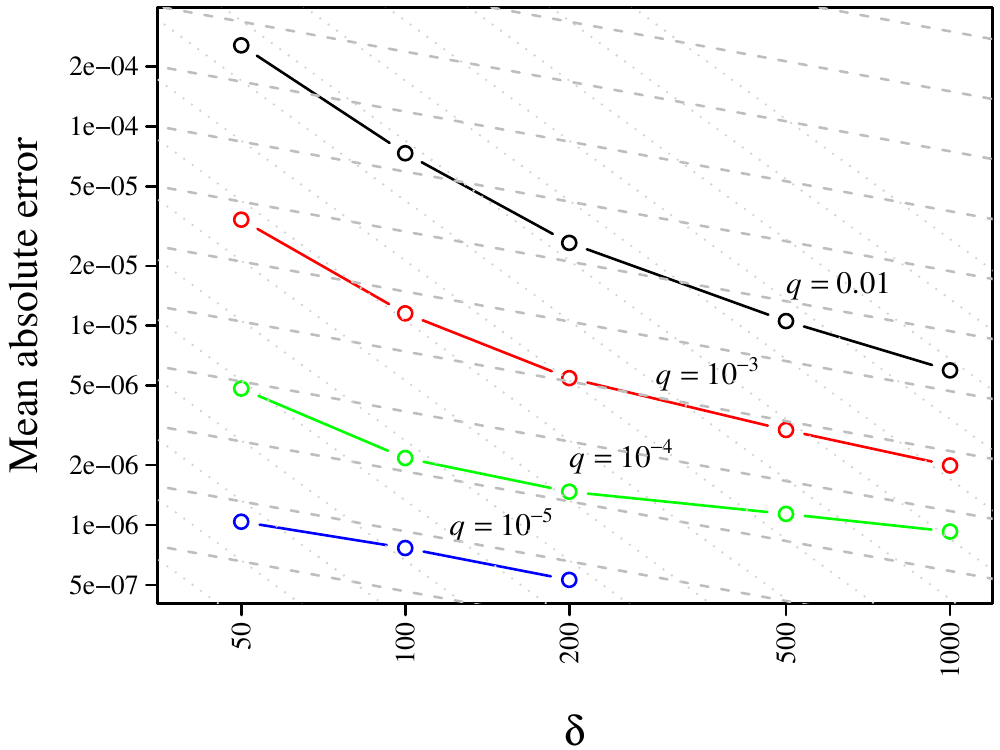} 
   \caption{The scaling of quantile estimation for various values of compression factor $\delta$ and $q$. 
   The general pattern is that absolute error scales like $1/\delta^2$ for small values of $\delta$ and like $1/\sqrt{\delta}$ for values larger than some cutoff. The cutoff is higher for larger values of $q$ (roughly 500 for $q=0.01$ but only 100 for $q=10^{-5}$). The grey dashed lines provide a reference for $1/\sqrt{\delta}$ scaling, the dotted grey lines show $1/\delta^2$.   At $q=10^{-5}$ error goes to zero for values $\delta > 200$ and cannot be shown on a log scale. The same happens for $q=10^{-6}$ except error is zero for all values of $\delta$. The data shown here are for $k_2$, but the results for $k_3$ are nearly indistinguishable.}
   \label{fig:accuracy-scaling}
\end{figure}

Not surprisingly, there is a trade-off between the size of the $t$-digest as controlled by the compression parameter $\delta$ versus the accuracy which which quantiles can be estimated.  Figure \ref{fig:accuracy-scaling} shows the trade-off for different quantiles. Of particular interest is the fact that for smaller values of $\delta$, errors appear to decrease roughly with the  square of the number of retained centroids (or equivalently, the compression parameter $\delta$) and with the square root for larger values of $\delta$. The original (now known to be incorrect, or at least incomplete) intuition underlying the $t$-digest suggested that errors would decrease with the square of size of the clusters in the digest as would be the case of a step-wise approximation of a continuous curve as step size is decreased. In fact, numerical experiments suggest the dominant sources of error in $t$-digest are the difference between the mean and median of each cluster and deviation of the actual distribution of samples in a cluster from the linearly interpolation. Both of these error sources should scale with the inverse square root of the number of samples in a cluster.

In particular, since cluster size for a particular value of $q$ should be directly proportional to $\delta$,  the observed scaling of error with $\delta$ appears to contradict the original intuition and support square root scaling of errors, at least for larger values of $\delta$. This discrepancy explains why $k_1$ does not perform as well as $k_2$ and $k_3$ in terms of relative error, but a rigorous explanation of the scaling of error with $\delta$ is still not available. In practical terms, however, the $t$-digest clearly performs well.

\subsection{Computing $t$-digests in parallel}
With large scale computations, it is important to be able to compute aggregates like the $t$-digest on separate portions of the input in parallel and then combine those aggregates.  

For example, in a map-reduce framework such as Hadoop, a combiner function can build a $t$-digest on the outputs of several mappers and then a single reducer can be used to combine the outputs of the combiners to compute the $t$-digest for the entire data set.  

Another example can be found in certain databases such as Apache Druid which maintains tree structured indices and allows the programmer to specify that particular aggregates of the data being stored can be kept at interior nodes of the index.  The benefit of this is that aggregation can be done almost instantaneously over any contiguous sub-range of the index.  The cost is quite modest with only a $O(\log(N))$ total increase in effort over keeping a running aggregate of all data where $N$ is the depth of the tree.  In many practical cases, the tree can be no more than two or three levels and still provide fast enough response.  For instance, if it is desired to be able to compute quantiles for any period up to 30 years in 30 second increments, simply keeping $t$-digests for intervals of 30 seconds, 2.5 hours and 31 days is likely to be satisfactory because at most about 300 digests will ever need to be merged.  

The most straightforward to merge $t$-digests is to use Algorithm \ref{alg:merge} which sorts all centroids from the $t$-digests being merged and then consolidates them as much as possible while maintaining the $t$-digest size criterion. As with non-parallel implementations, stratified merging is useful where a higher value of $\delta$ is used when data is inserted into the smaller digests and a smaller value of $\delta$ is used when combining the digests. 

\begin{figure}[htb] 
   \centering
   \includegraphics[width=5.5in]{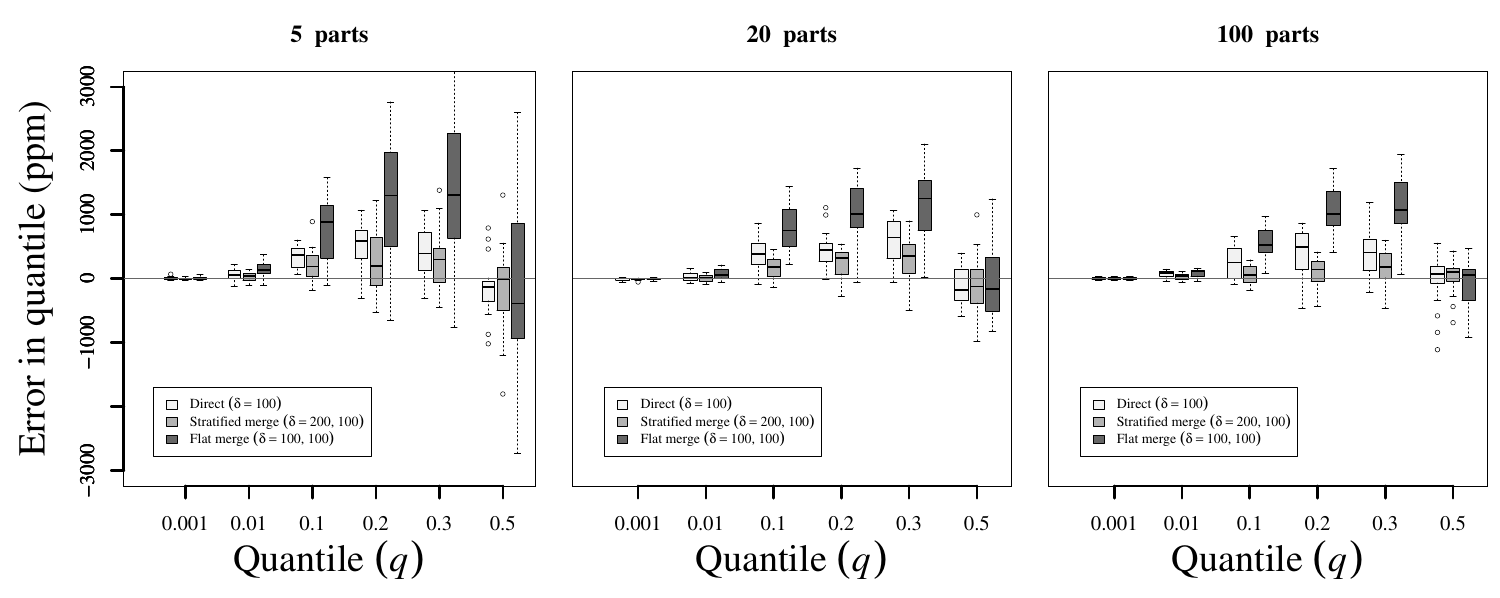} 
   \caption{Accuracy of a $t$-digest accumulated directly is nearly the same as when the digest is computed by combining digests from 5, 20 or 100 equal sized paritions of the data.  All panels were computed by 20 repetitions of aggregating 1,000,000 values. }
   \label{fig:merge}
\end{figure}

Figure \ref{fig:merge} illustrates how this turns out. This figure shows accuracy achieved when summarizing 1,000,000 samples across 20 trials. The panels, taken from left to right, show the results of merging 5, 20 and 100 sub-digests. Within each panel, the three bars represent the use of a single digest which internally uses $\delta=300,100$ stratified merging, merged digests with $\delta=200$ in the sub-digests and $\delta=100$ in the result and non-stratified merge with $\delta=100$. At all levels of parallelism, the non-stratified merge result was worse than the non-merged stratified result. Stratified parallel merging was comparable to the non-parallel merge, though a bit more variable, with 5-way parallelism, but got progressively more accurate with high levels of parallelism.

In terms of speed, merging $t$-digests is relatively cheap. In one experiment, 1000 digests were used to summarize 1,000,000 samples each and these digests were then merged into a single digest. It took about 150ms to fill each digest and about 150 ms to merge all thousand digests. Each digest is about 1kB in this example so a total of about 1MB of digests would need to be moved to the merge. Overall, this means that, with sufficient hardware (roughly 1000 cores), a $t$-digest can be produced from a billion data points in less than half a second. 

\subsection{Comparison with Q-digest}
\begin{figure}[htb] 
   \centering
   \includegraphics[width=4.5in]{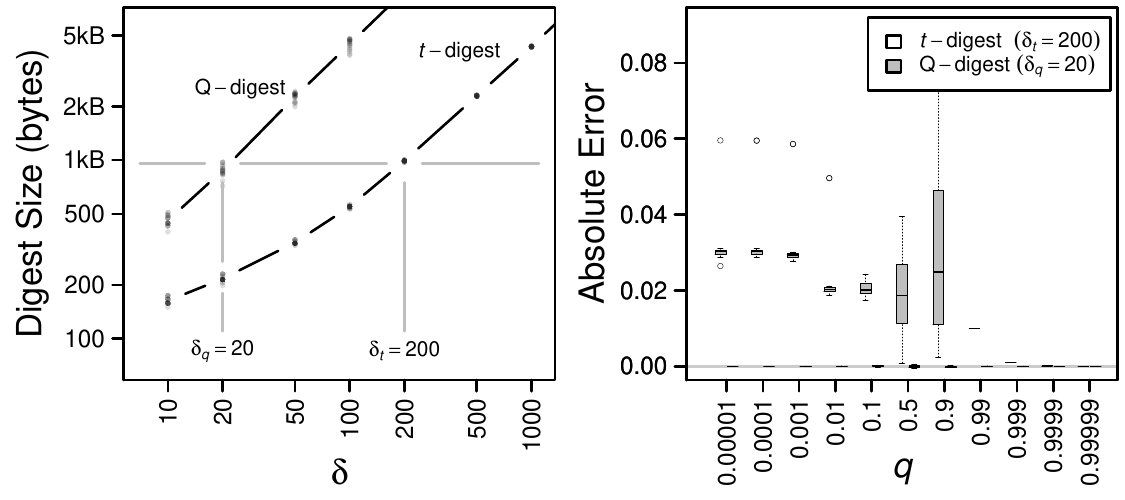} 
   \caption{The left panel shows the size of a serialized Q-digest and $t$-digest versus compression parameter $\delta$ for 100,000 uniformly distributed samples. Trials were repeated 20 times. The left panel shows how compression parameters $\delta_q$ and $\delta_t$ are chosen to control digest size to approximately 1k byte. The right panel shows absolute error for various values of $q$.  With this vertical scale, $t$-digest errors are only barely distinguishable from zero near the median.  }
   \label{fig:qd-comparison}
\end{figure}

In terms of error size, the 4.0 preview implementation of $t$-digest dominates the implementation of the Q-digest\citep{qdigest} from the popular stream-lib package \citep{github:stream} when digests of the same size are compared.  This is shown in Figure \ref{fig:qd-comparison}.  In the left panel, the relationship between the compression parameter $q$ and size is shown for both digests for 20 trials where 100,000 uniformly distributed samples are inserted into each kind of digest.  To force the digests to be equal in size compression parameter values of $\delta_q=20$ for Q-digest  and $\delta_t = 200$ for $t$-digest were chosen. This gives an average digest size of about $874$ bytes for Q-digest and 990 bytes for $t$-digest, as shown in the left panel. The panel on the right compares errors under these conditions and shows that the $t$-digest errors are more than an order of magnitude smaller for all values of $q$ (actually more than two orders of magnitude better). For values near $q=0$ the $t$-digest produces absolute errors 3-5 orders of magnitude smaller than the Q-digest does. When compared on distributions with high amounts of skew, the results are even more striking in favor of the $t$-digest.
Note that the parameters chosen for this comparison depend on the number of samples because the size of the Q-digest increases roughly logarithmically with the number of samples while the size of the $t$-digest is strictly bounded and is nearly constant once the number of samples exceeds $2 \delta$. 

\section{Conclusion}
The $t$-digest is a novel on-line algorithm that dominates the previously state-of-the-art Q-digest in terms of accuracy and size.  The $t$-digest provides accurate on-line estimates of a variety of of rank-based statistics including quantiles and trimmed mean.  The core ideas of the algorithm are simple and empirically demonstrate high accuracy.  The $t$-digest can also be used in parallel applications or in OLAP-style indexes.  

The $t$-digest algorithm is available in the form of an open source, well-tested implementation maintained the author \citep{t-digest-project}.  It has already been adopted by large companies such as Netflix and Microsoft for internal monitoring and has been included in open source libraries such as Apache Mahout and stream-lib and has been incorporated into prominent open-source projects such as Apache Lucene and Elasticsearch.

\bibliographystyle{authordate1}
\bibliography{refs}{}

\end{document}